\documentclass[letter]{aa}  

\usepackage{graphicx}
\usepackage{txfonts}
\usepackage[colorlinks,allcolors=blue]{hyperref}
\newcommand{\kmpers}      {\mbox{\rm km~s$^{-1}$}}
\newcommand{\kms} {\kmpers}

\begin{document}

   \title{OGHReS: Large-scale filaments in the outer Galaxy}

   \author{D. Colombo\inst{1}\thanks{dcolombo@mpifr-bonn.mpg.de},
          C. K\"onig\inst{1},
          J. S. Urquhart\inst{2},
          F. Wyrowski\inst{1},
          M. Mattern\inst{3},
          K. M. Menten\inst{1},
          M.-Y. Lee\inst{4},\\
          J. Brand\inst{5},
          M. Wienen\inst{1},
          P. Mazumdar\inst{1},
          F. Schuller\inst{1,6},
          S. Leurini\inst{7}
        }

   \institute{Max-Planck-Institut f\"ur Radioastronomie, Auf dem H\"ugel 69, 53121 Bonn, Germany
   \and
   Centre for Astrophysics and Planetary Science, University of Kent, Canterbury, CT2\,7NH, UK
   \and
   Commissariat à l’énergie atomique et aux énergies alternatives, 91191 Gif-sur-Yvette, Saclay, France
   \and
   Korea Astronomy and Space Science Institute, 776 Daedeok-daero, Yuseong-gu, Daejeon 34055, Republic of Korea
   \and
   INAF - Istituto di Radioastronomia \& Italian ALMA Regional Centre, via P. Gobetti 101, 40129, Bologna, Italy
   \and
   Leibniz-Institut für Astrophysik Potsdam (AIP), An der Sternwarte 16, D-14482 Potsdam, Germany
   \and
   INAF - Osservatorio Astronomico di Cagliari, Via della Scienza 5, 09047 Selargius (CA), Italy
   }

   \date{Received XXX; accepted XXX}

 
  \abstract
    {Filaments are a ubiquitous morphological feature of the molecular interstellar medium and are identified as sites of star formation. In recent years, more than 100 large-scale filaments (with a length $>10$\,pc) have been observed in the inner Milky Way. As they appear linked to Galactic dynamics, studying those structures represents an opportunity to link kiloparsec-scale phenomena to the physics of star formation, which operates on much smaller scales. In this letter, we use newly acquired Outer Galaxy High Resolution Survey (OGHReS) $^{12}$CO(2-1) data to demonstrate that a significant number of large-scale filaments are present in the outer Galaxy as well. The 37 filaments identified appear tightly associated with inter-arm regions. In addition, their masses and linear masses are, on average, one order of magnitude lower than similar-sized molecular filaments located in the inner Galaxy, showing that Milky Way dynamics is able to create very elongated features in spite of the lower gas supply in the Galactic outskirts.}


   \titlerunning{OGHReS: outer Galaxy large-scale filaments}

   \authorrunning{D. Colombo, C. K\"onig, J. Urquhart et al.}

   \maketitle

\section{Introduction}\label{S:introduction}
Since the early observations with \emph{Herschel} \citep{andre10}, it became evident that the molecular interstellar medium (ISM) has a preference to organise itself in filaments. Filaments are ubiquitous in the Galaxy and are found on a large variety of scales. Networks of infrared dust extinction filamentary features (indication high column densities) constitute molecular clouds \citep[e.g. ][]{andre10,menshchikov2010,arzoumanian2011,schneider2012,li2016,schisano2020}. Elongated structures within clouds are also observed in emission \citep{panopoulou2014,suri2019}. Clouds themselves are mostly elongated structures \citep[][Neralwar et al. in prep.]{duarte-cabral2021}. Additionally, bundles of smaller scale `fibers' have been found within those filaments \citep{hacar2013,hacar2016,henshaw2016}. Extremely long (even longer than 100\,pc) and velocity coherent large-scale filaments (LSFs) have been observed in the Milky Way as part of spiral arms or inter-arm regions \citep{jackson2010,goodman2014,ragan2014,zucker2015,abreu-vicente2016,li2016,du2017,zucker2018,lin2020,schisano2020}. As the physics of filaments appears tightly connected to star formation \citep[e.g. ][]{andre10,elia2013,andre2014}, a great deal of studies have been put forward to explain the origin and evolution of those structures. Filaments have been linked to ISM turbulence \citep{padoan2001}, and also to cloud–cloud collisions, expanding shells, and magnetic fields \citep[e.g ][]{hartmann_burkert2007,heitsch08,molinari2014}. LSFs instead appear to be formed by large-scale phenomena such as spiral arm shocks, galactic shear, and supernovae (SNe) feedback. 
Simulations have found LSFs in both spiral arms and inter-arm regions \citep[e.g. ][]{dobbs_pringle2013,smith2014,smith2020}. However, the simulations of \cite{smith2014} predicted that LSFs originate within spiral arms, supporting the Galactic `bone' paradigm \citep{goodman2014,zucker2015}, while the work by \cite{duarte-cabral2016}, whose simulations also include SNe feedback and gas self-gravity, indicate that LSFs originate exclusively by the intense shear of the inter-arm regions. 
Upon entry in the spiral arms, those filaments lose their elongated morphology and merge to become large cloud complexes \citep{duarte-cabral2017}. More recently, the work of \cite{smith2020}, instead, simulated LSFs in both spiral arms and inter-arm regions. They also found that those structures tend to be longer than filaments that formed under the action of SNe feedback, which in turn have large velocity gradients and might be transient features of the ISM.

In this letter we continue the search of LSFs in the outskirts of the Milky Way. The outer Galaxy presents different characteristics compared to the inner Galaxy, in terms of the radiation field, metallicity, H\textsc{i} density, and gas-to-dust ratio values (see Appendix~\ref{A:outer_galaxy} for more details). Therefore, it represents an opportunity to study the properties of the LSFs in largely unexplored environmental conditions. Using $^{12}$CO(2-1) data from the OGHReS science demonstration phase (SD1; see Appendix~\ref{AA:oghres} and K\"{o}nig et al., in prep. for further details), we identified tens of outer Galaxy LSFs (hereafter OGLSFs) and compared their properties with similar-scale objects observed in the inner Milky Way.

\begin{figure*}
\centering
\includegraphics[width=1\textwidth]{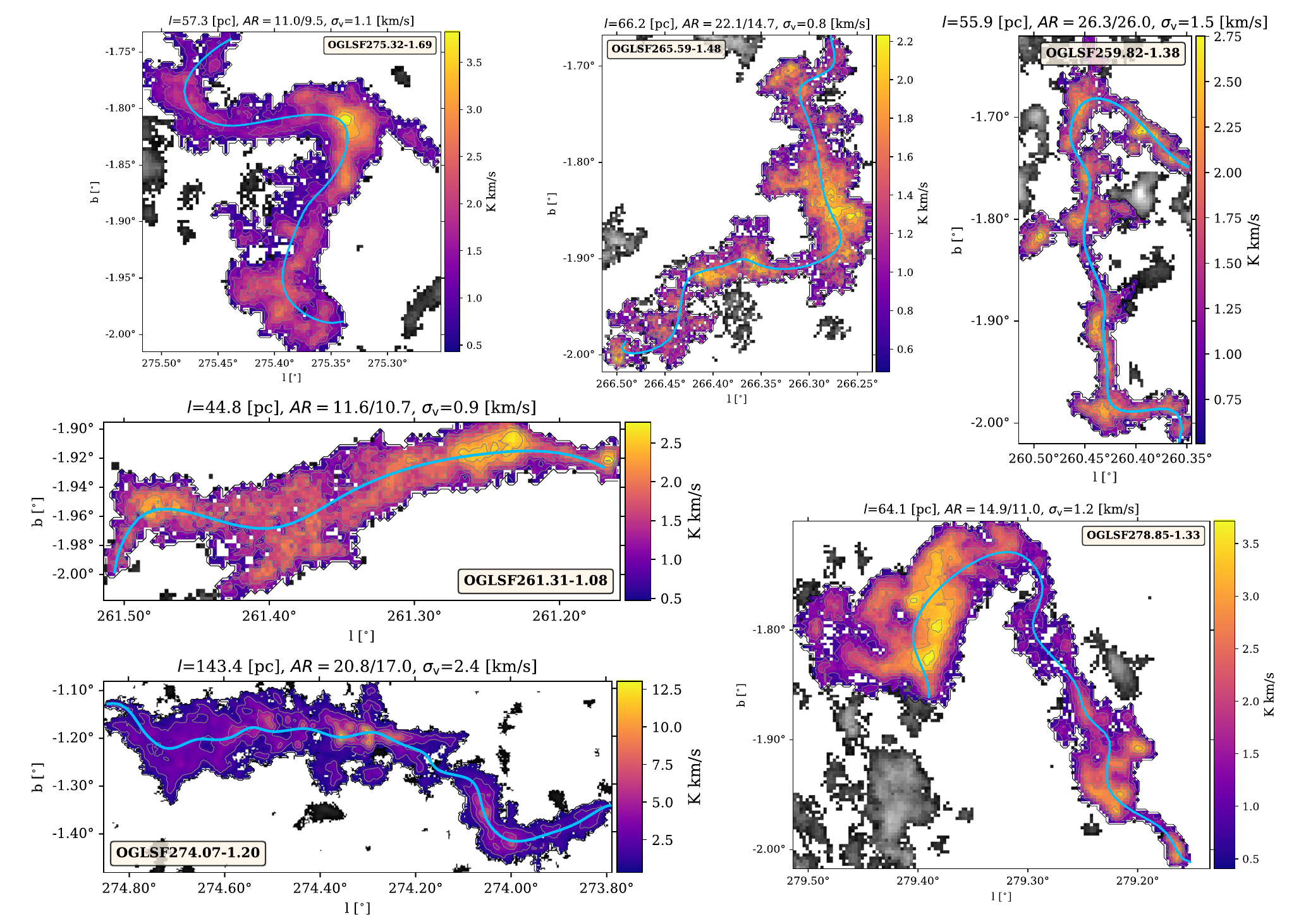}
\caption{Integrated intensity maps of the $^{12}$CO(2-1) emission from some of the most prominent outer Galaxy LSFs identified (in colour) within the structure masked data defined by the dendrogram. In grey-scale, the emission across the line-of-sight that does not belong to the structure is shown. The cyan line displays the filament spine. In the box, the structure name is indicated. In the panel titles, the length ($l$), the aspect ratio ($AR=$\emph{length/width}, with the width inferred from the medial axis method/RadFil), and the velocity dispersion ($\sigma_{\rm v}$, calculated as the intensity-weighted second moment of velocity) of the structures are shown. The velocity range spanned by the structures is reported in Table~\ref{T:catalogue} as 5$th$ and 95$th$ percentiles of the LSF velocity distribution.}
\label{F:oglsf_mom0}
\end{figure*}

\begin{figure*}
\centering
\includegraphics[width=1\textwidth]{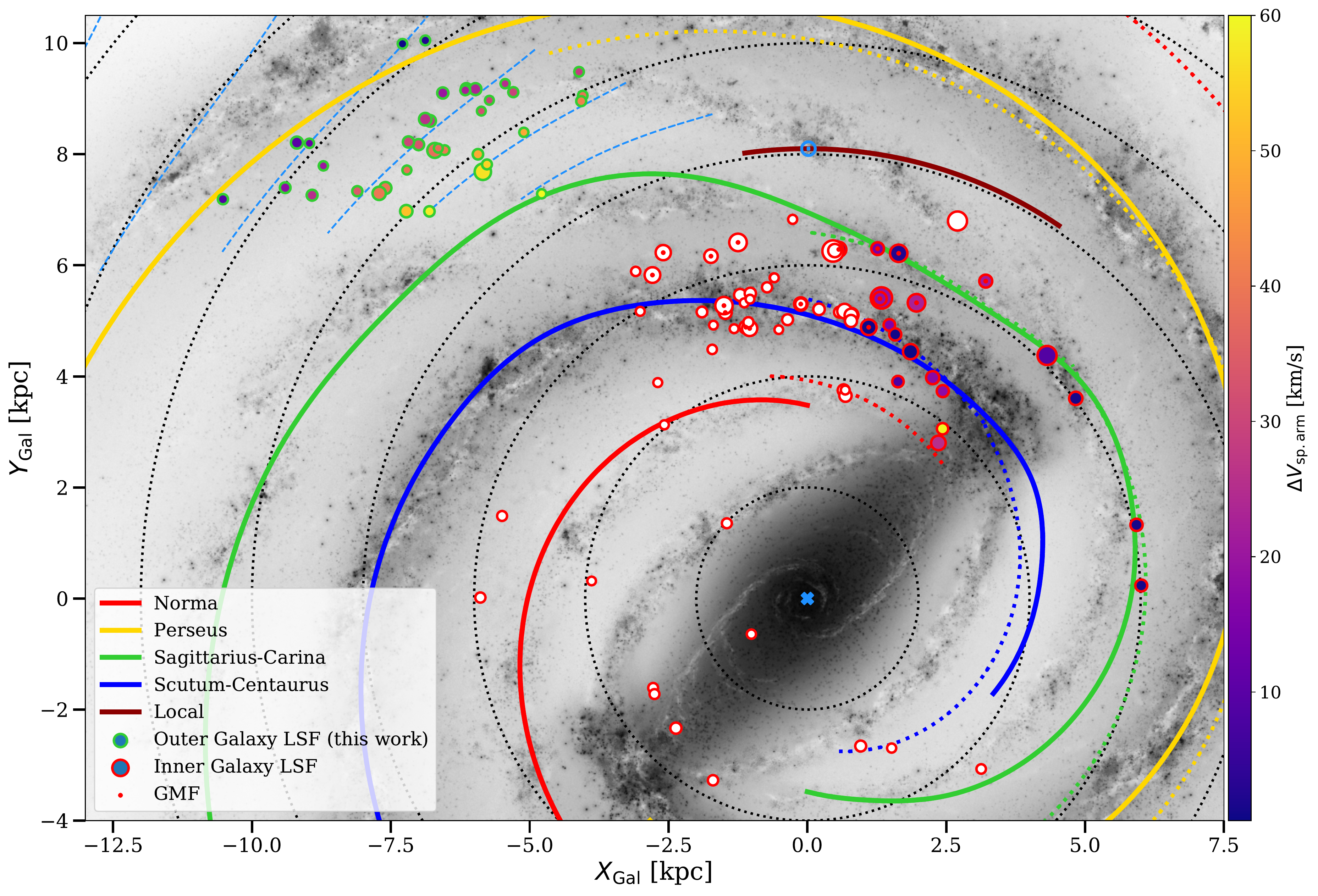}
\caption{Location of the large-scale filaments (LSFs) across the Milky Way disc. Inner Galaxy LSFs and outer
Galaxy LSFs are indicated with red and green contoured circles, respectively. The size of the circles is proportional to the filament lengths. The inner colours of markers show the velocity offset ($\Delta V_{\rm sp. arm}$) with respect to the closest spiral arm (see the colour scale on the right). White-filled markers are for inner Galaxy LSFs where this quantity is not provided by their original catalogues. Inner Galaxy giant molecular filaments (GMFs) are highlighted with red dots.  Coloured full lines mark the position of the spiral arms following \cite{taylor_cordes1993} models, as adapted in \cite{urquhart2021}. The `Local' arm (or spur) parameters are taken from \cite{reid2019}. Dotted coloured lines show the correspondent spiral arm loci from \cite{reid2016}, used to define the spiral arm associations in the inner Galaxy. Galactocentric circles drawn with black dotted lines are spaced 2\,kpc apart. Cyan dashed arcs show loci with equal velocity between 20-120\,\kms{} (from the Galactic centre outwards) and spaced 20\,\kms{} apart. The cyan `X' indicates the Galactic centre and `$\odot$' the position of Sun. The figure was produced using the {\sc mw-plot} package (available at \url{https://milkyway-plot.readthedocs.io/en/latest/}).}
\label{F:mw_faceon}
\end{figure*}

\section{Results}\label{S:results}
\begin{figure*}
\centering
\includegraphics[width=1\textwidth]{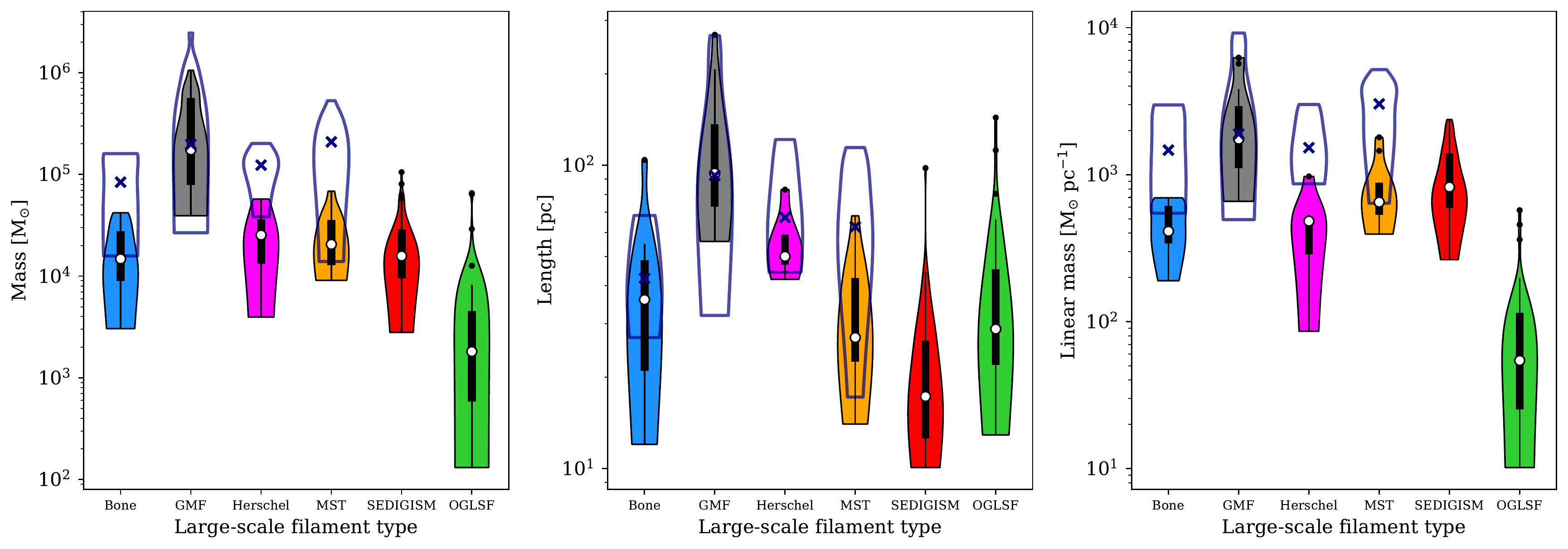}
\caption{Violin- (colours) and box-plot (black) representations of the distributions of mass (left), length (middle), and linear mass (right) of Galactic LSFs. Following \cite{zucker2018}, IGLSFs are divided into Milky Way `bones' (blue), GMFs (grey), Herschel filaments (magenta), minimum spanning tree (MST) filaments (yellow). IGLSFs identified from the Structure, Excitation, and Dynamics of the Inner Galactic InterStellar Medium (SEDIGISM, \citealt{schuller2017}) survey data are in red (from \citealt{mattern2018b}). OGLSFs, observed within our data, are in green. More information on IGLSFs are reported in Appendix~\ref{AA:ancillary}. The white circles show the median for each property and sub-sample. For comparison, the transparent violins show the measurements of \citealt{zhang2019} for the same category of structures, where the `x' indicates the median of the samples.}
\label{F:props}
\end{figure*}

To search for filaments in OGHReS data (between $250<l<280$ deg), we masked the data considering only the high signal-to-noise ratio (S/N) emission and we used a dendrogram analysis \citep{rosolowsky08} to isolate coherent structures (see Appendix~\ref{AA:method} for more details).  As OGLSFs, we considered all of the `trunks' of the dendrogram with a length $>10$\,pc, which is a clear filamentary morphology (or a typical aspect ration $AR>10$), and observed at $V_{\rm LSR}>40$\,km\,s$^{-1}$. Lengths and $AR$ were measured through the `medial axis analysis' and RadFil \citep{zucker2018}. Using this method, we identified 37 OGLSFs, which is over a factor of 7 more than the LSFs currently identified in the outer Galaxy \citep{du2017}. However, while we focused on all filamentary structures with a length $>10$\,pc, the work of \cite{du2017} considers only the longest structures (with a length $>55$\,pc). In this aspect, we have a comparable number of filaments (ten objects with a length $\sim40-144$\,pc; see Fig.~\ref{F:oglsf_mom0} and Appendix~\ref{A:catalog}), despite having searched for them in roughly a third of the Galactic area surveyed by \cite{du2017}. Additionally, our search focuses on only the most prominent structures and does not consider all LSFs present in OGHReS SD1 data as we have included only connected and isolated structures at $V_{\rm LSR}>40$\,km\,s$^{-1}$. Indeed, we have also observed LSFs within the `local' gas (at $V_{\rm LSR}\sim0$\,km\,s$^{-1}$), objects with a length and $AR$ below our assumed thresholds, and velocity coherent fragments of filaments that dendrograms alone are not able to merge into a single structure. The 37 OGLSFs represent a lower limit of the actual number of filaments present in the survey. We might expect to find at least $\sim120$ similar properties of LSFs in the full 100 deg$^2$ span by the full OGHReS data, once the survey is completed. A full catalogue of all filamentary structures present within OGHReS data will be presented in a future work. This would be comparable to the current number of LSFs observed in the inner Galaxy (approximately 70 objects), whose properties we use here for comparison. Those inner Galaxy large-scale filaments (hereafter IGLSFs) have been found by different studies and have been named according to their appearance or identification methods (see Appendix~\ref{AA:ancillary} for more details). The properties of the IGLSFs that we use here have been re-elaborated by \cite{zucker2018} and \cite{zhang2019}, with the addition of the filaments from \cite{mattern2018a}.

Integrated intensity maps of the $^{12}$CO(2-1) emission from the longest OGLSFs (with a length $>40$\,pc) are compiled in Fig~\ref{F:oglsf_mom0}, which illustrates the complexity of the LSFs observed. All selected LSFs are velocity coherent structures (where the median of the velocity dispersion distribution is $\sim 1$\,km\,s$^{-1}$) which might indicate that those filaments are young or quiescent.

The location of the OGLSFs across the Milky Way disc is shown in Fig.~\ref{F:mw_faceon}. IGLSFs are observed at heliocentric distances $1.3-11.6$\,kpc, similar to OGLSFs that show distances $4.2-10.6$\,kpc. Nevertheless, IGLSFs are distributed within $R_{\rm Gal}=1-7$\,kpc, while our OGLSFs are further away from the Galactic centre and show $R_{\rm Gal}=9-13$\,kpc. Despite the giant molecular filaments (GMFs) class, for which two-thirds of the objects are found in the inter-arm region \citep{zucker2018} (see also \citealt{ragan2014,abreu-vicente2016}), other IGLSF categories appear closer to or even within spiral arms \citep{jackson2010,goodman2014,zucker2015,wang2015,wang2016,zucker2018}. Instead, OGLSFs seem to have all inter-arm features, as only three of them have a velocity offset ($\Delta V$) with respect to the closest spiral arm (the Perseus arm, assuming the \citealt{taylor_cordes1993} spiral arm models, as adopted in \citealt{urquhart2021}) $<10$\,km\,s$^{-1}$. The others are all kinematically far from the Perseus arm with a median $\Delta V=30.2$\,km\,s$^{-1}$ (and a median absolute deviation of 8.7\,km\,s$^{-1}$; see also Table~\ref{T:catalogue}). 

A comparison of some of the properties of the LSFs observed in the inner and outer Galaxy is shown in Fig.~\ref{F:props}. 
Here, we only compare properties that are also available from the SEDIGISM catalogue. The OGLSFs show some of the smallest values of mass within the full sample, ranging from $\sim10^2-7\times10^4$\,M$_{\odot}$, while most of the inner Galaxy structures show masses $\sim3\times10^3-5\times10^5$\,M$_{\odot}$. An exception are the GMFs with masses of  $\sim2.5\times10^4-2.5\times10^6$\,M$_{\odot}$. By construction, all LSFs have a minimum length $\sim10$\,pc. However, Herschel filaments show a minimum length of $42$\,pc. The GMFs are the longest filaments, on average, in the sample having lengths from $56-270$\,pc, with a median of $94$\,pc (for \citealt{zucker2018} estimations, while considering \citealt{zhang2019} measurements, GMFs show a minimum length $\sim31$\,pc). Indeed, only the longest structures in the other categories reach the median length of the GMFs. Due to their low mass content, the OGLSFs show some of the lowest median values of linear mass (mass over the length of the structures) between the full sample, having $\sim10-600$\,M$_{\odot}$\,pc$^{-1}$, while for the IGLSFs we observe linear masses $\sim86-9200$\,M$_{\odot}$\,pc$^{-1}$, where the largest linear mass is again observed for the GMFs (for more details regarding the mass estimation reliability across the different works, see Appendix~\ref{A:mass}). 

\section{Discussion and conclusions}\label{S:conclusions}
In the previous section, we observed several properties that distinguish the OGLSFs from the IGLSFs. In particular, many IGLSFs are associated with the spiral arms, while all the OGLSFs appear as inter-arm objects. In addition to having similar lengths as the LSFs in the inner Galaxy (whose lower limit is set by construction), OGLSFs are in general less massive, resulting in much lower linear masses compared to IGLSFs. Taken together, these might indicate a different origin, and possibly evolution, for the filaments observed in the spiral arms versus inter-arm regions, and inner versus outer Galaxy. The analysis of \cite{zucker2019} performed on \cite{smith2020} simulations showed that spiral arm filaments tend to have masses and linear masses up to $5\times$ higher compared to similar-length inter-arm objects, indicating that the Galactic environment can have a significant impact on their properties. Even if this work is based on very long filaments ($>100$\,pc), its conclusions are in line with our results. Nevertheless, from an observational point of view, some caveats need to be considered. First of all, the exact morphology of the Milky Way is not well known \citep{dobbs_baba2014}, especially in its outskirts \citep{koo2017}, which renders the exact localisation of the filaments with respect to the spiral structure challenging. Indeed, only a third of the IGLSFs analysed by \cite{zucker2018} can be spatially and kinematically associated with spiral arms. However, this result does not include all objects in \cite{zucker2018}, and it is model-dependent \citep[e.g. ][]{zucker2015}. Additionally, distance measurements in the inner Galaxy are challenging due to a number of effects, such as the kinematic distance ambiguity (KDA) and spiral arm velocity crowding. The outer Galaxy does not suffer from these issues, therefore kinematic distance-dependent properties of the OGLSFs can be considered to be more robust than for their inner Milky Way counterparts. Nevertheless, for sources observable from the  northern hemisphere, trigonometric parallax distances determined for radio wavelength maser sources with very long baseline interferometry (VLBI) are used to `anchor' the locations and dimensions of spiral arms in the first and second Galactic quadrants \citep[and part of the third][]{reid2019}. So far, the distances of only very few southern hemisphere have been measured with VLBI \citep{Krishnan2015, Krishnan2017} and none of them are within the longitude range considered here ($250<l<280$ deg). 

It is interesting to note that (considering the analysis of \citealt{zucker2018}) two-thirds of the GMFs appear to be inter-arm features, as our OGLSFs. However, the property distributions of GMFs and OGLSFs are the most discrepant within the sample. Some of these differences might be intrinsically methodological. For example, the GMF category contains the longest objects in the sample, while only a few OGLSFs reach the length of an average GMF. GMFs have been firstly identified in extinction using GLIMPSE data \citep{benjamin2003}; subsequently, their velocity coherence has been verified with GRS $^{13}$CO(1-0) data \citep{jackson2006}. We used the GMF properties presented in \cite{zucker2018} and \citealt{zhang2019}. In our case, instead, all of those operations were made via emission data. Our searching method does not allow us to connect fragments of a filament, but only finds velocity coherent regions that, as a whole, stand out above the noise level. Those velocity coherent fragments might be abundant in our data given the low brightness of the gas emission in the outer Galaxy and the limited sensitivity. Therefore our catalogue might be biased towards shorter filaments. 
Considering that various methods appear to provide similar mass estimates for IGLSFs and OGLSFs (see Appendix~\ref{A:mass}), mass differences between GMFs (or generally also of the IGLSFs) and OGLSFs 
might be due to less molecular gas supply in the outer Galaxy compared to the inner Galaxy. The molecular gas distribution across the disc of the Milky Way varies significantly with the Galactocentric radius. The molecular gas mass surface density decreases by a factor of a few moving from the inner to the outer Galaxy (see \citealt{heyer_dame2015}, their Fig.~7). For instance, for similar large-scale effects operating in the inter-arm regions across the Milky Way, there might simply not be enough gas available to build up filaments as massive as GMFs (and with similar linear masses) in the outer Galaxy. Indeed, the OGLSFs observed here have a peculiar filamentary morphology, but their masses are in-line with the typical mass of clouds in the outer Galaxy ($<10^4$\,M$_{\odot}$, \citealt{heyer01}, but see also \citealt{miville_deschenes2017}), whose maximum value appears to be an order of magnitude lower compared to the clouds in the Solar circle (see \citealt{brand_wouterloot1995}, their Fig.~8). The rapid decrement of the molecular gas mass surface density with the distance might also be the reason why our OGLSFs are almost all observed in the inter-arm region. At the longitudes spanned by our data, the lines-of-sight cross two spiral arms, the Perseus and the Outer arms. However, as they are both located at far distances ($R_{\rm Gal}>12$\,kpc), they appear faint and patchy in our data, mostly encompassing low S/N flux (see K\"onig et al. in prep.). Therefore, OGHReS SD1 imaged mostly emission originating from the inter-arm region.
Only the longest (143.4\,pc) filament observed within our sample, OGLSF274.07-1.20 (see Fig.~\ref{F:oglsf_mom0}), has properties comparable with inner Galaxy filaments. Additionally, similar to the IGLSFs \citep[e.g. ][]{zhang2019}, it is actively forming stars and embedding a SN remnant \citep{duncan1996} and several HII regions \citep[e.g. ][]{haverkorn2006,culverhouse2011}. As SNe feedback appears to be the main mechanism responsible for filament dissipation \citep[see simulations by][]{smith2020}, this structure might be transient in nature.

In summary, we have assembled the largest catalogue, to date, of emission-selected outer Galaxy large-scale filaments (with a length $\gtrapprox10$\,pc and $AR\gtrapprox10$), consisting of 37 velocity coherent objects located at $R_{\rm Gal}>9$\,kpc. Almost all OGLSFs appear to be part of inter-arm regions and have masses and linear masses, on average, one order of magnitude lower than other LSFs observed in the inner Galaxy. The completion of the OGHReS project, with 70\,deg$^2$ more molecular gas emission mapped, will provide access to more than 100 similar objects in the outer Milky Way and shed new light on how galactic dynamics shapes molecular clouds (and ultimately star formation) in a largely unexplored environment of our own Galaxy.

\begin{acknowledgements}
      The authors thank the anonymous referee for the feedback that helped improving the clearness of the paper. DC acknowledges support by the German
      \emph{Deut\-sche For\-schungs\-ge\-mein\-schaft, DFG\/} project
      number SFB956A. DC thanks Miaomiao Zhang for the help and the useful discussion regarding the inner Galaxy filament properties. SL acknowledges financial support from INAF through the grant Fondi mainstream `Heritage of the current revolution in star formation: the Star-forming filamentary Structures in our Galaxy' This research made use of Astropy,\footnote{http://www.astropy.org} a community-developed core Python package for Astronomy \citep{astropy2013, astropy2018}; matplotlib \citep{matplotlib2007}; numpy and scipy \citep{scipy2020}.

\end{acknowledgements}

\bibliographystyle{aa}
\bibliography{cold}

\appendix

\section{A glimpse into the outer Galaxy environment}\label{A:outer_galaxy}
In addition to the decrements of molecular gas surface density and the molecular cloud mass across the Galactic disc (see Section~\ref{S:conclusions}), several studies have shown that many other parameters display gradients with respect to the Galactocentric radius, $R_{\rm Gal}$, making the outer Galaxy an environment profoundly different compared to the inner Milky Way. For example, \cite{giannetti2017} measured a gas-to-dust, $\gamma$, gradient $\log(\gamma)=0.087R_{\rm Gal}+1.44$. Such a gradient would imply a large increment of the gas-to-dust ratio from the inner to the outer Galaxy with, for example, $\gamma(R_{\rm Gal}=5\,\mathrm{kpc})\sim75$ and $\gamma(R_{\rm Gal}=15\,\mathrm{kpc})\sim556$. At the same time, they inferred the variation of gas metallicity, $Z$, with respect to $R_{\rm Gal}$: $\log(Z)=-0.056R_{\rm Gal}-1.176$. Considering this, the metallicity at the Sun location ($R_{\rm Gal}=8.1$\,kpc) is $\sim2.5$ higher than the metallicity at $R_{\rm Gal}=15$\,kpc. Various works \citep[see][and references therein]{kalberla_kerp2009} have reported decrements of the H\textsc{i} surface density $\Sigma_{\rm HI}=30\exp((R_{\rm Gal}-R_0)/3.75)$\,M$_{\odot}$\,pc$^{-2}$. This quantity, however, appears to saturate at $\Sigma_{\rm HI}\sim10$\,M$_{\odot}$\,pc$^{-2}$ up to $R_{\rm Gal}=15$\,kpc. A reduced radiation field has also been measured in the outer Galaxy. For example, as the $\gamma-$ray emissivity appears to be directly correlated with the H\textsc{i} column density, \cite{bloemen1984} found that $\gamma-$ray emissivity decreases by $15\%$ for $R_{\rm Gal}>R_0$ with only $25\%$ of the $\gamma-$ray intensity coming from $14<R_{\rm Gal}<17$\,kpc.

\section{Data and methods of analysis}\label{A:data_method}

\subsection{OGHReS science demonstration data}\label{AA:oghres}

OGHReS is a large programme that is mapping the $^{12}$CO(2-1), $^{13}$CO(2-1), and C$^{18}$O(2-1) emission across $180\leq l \leq 280$\,deg using the PI230 and nFLASH230 receivers mounted on the Atacama Pathfinder EXperiment \citep[APEX, ][]{guesten2006}. The latitude coverage varies and follows the Galactic warp, covering $-2.0 < b < -1.0$\,deg in the range of $280 > \ell > 225$\,deg then increasing in latitude up to $\ell=205$\,deg from where it covers $-0.5 < b < 0.5$\,deg down to $\ell=180$\,deg. Data present an angular resolution HPBW=$27^{\prime\prime}$, a spectral resolution of 0.25\,\kms{}, and a median $\mathbf{\sigma_{\rm RMS}=0.23\pm0.08}$\,K (T$_{\rm mb}$). Here, we use data from the science demonstration phase of the survey (OGHReS SD1), covering $250\leq l \leq 280$\,deg (see K\"onig et al. in prep. for more details).

\subsection{Inner Galaxy large-scale filament data}\label{AA:ancillary}

In the paper, we compare the location and properties of the OGLSFs considering the inner Galaxy LSFs (IGLSFs), which were collected and homogeneously reanalysed by \cite{zucker2018}, consisting of 45 objects. As in \cite{zucker2018}, here we adopted the same terminology to separate the IGLSFs as those structures that have been identified using different methods and that present slightly different properties (see Fig.~1 in \citealt{zucker2018} for further details). Those LSFs are named giant molecular filaments (GMFs, \citealt{ragan2014,abreu-vicente2016}), large-scale Herschel filaments \citep{wang2015}, Milky Way `bones' \citep{zucker2015}, and minimum spanning tree (MST) filaments \citep{wang2016}. To this sample, we added the SEDIGISM \citep{schuller2021} filaments assembled by \cite{mattern2018b}, starting from dust-continuum filaments originally extracted from ATLASGAL data \citep{schuller2009,li2016}. To be consistent with our sample that of \cite{zucker2018}, we considered only the fully correlated, single component filaments in the SEDIGISM catalogue that have a length $>10$\,pc, for a total of 30 objects. The final sample of IGLSFs consists of 75 objects. For consistency, we only considered objects whose coherency has been verified in emission. Therefore, here, we do not use larger filament samples, such as the Hi-GAL catalogue presented in \cite{schisano2020}.
For comparison, we also show the property measurements performed by \cite{zhang2019} by using the same LSFs considered by \cite{zucker2018} (e.g. excluding the ATLASGAL filaments from \citealt{zhang2019}, which have been homogeneously and fully reanalysed by \citealt{mattern2018b}).


\subsection{Outer Galaxy large-scale filament identification and property calculation}\label{AA:method}

To search for LSFs within OGHReS data, we used a procedure consisting of several steps. Following \cite{zucker2018}, we looked for coherent structures with a clear filamentary morphology (or an aspect ratio, $AR\gtrapprox 10$) and length $\gtrapprox 10$\,pc. In this work, we are only interested in `outer Galaxy' LSFs (or OGLSFs), therefore we generally only considered the emission with $V_{\rm LSR}>40$\,km\,s$^{-1}$.   

Firstly, we masked the data using a `dilate mask' technique \citep{rl06}, where we retained connected regions in the datacubes, whose pixels show a peak S/N > 2, but also contain a region with S/N>4. The S/N was calculated line-of-sight-wise and the noise map was obtained considering the standard deviation of the first and last ten line-free channels in the datacubes (which constitutes $\sim2\%$ of the total channels in the datacubes). A dendrogram analysis\footnote{Using {\sc astrodendro}, \url{https://dendrograms.readthedocs.io/en/stable/}} was performed in order to label the isolated regions in the masked data and to calculate their properties. In particular, the dendrogram produces three-dimensional (position-position-velocity) masks of the catalogued hierarchical structures, where each voxel that belongs to a given structure is labelled with the structure ID. The bi-dimensional projections of the structures on the plane of the sky were generated by flattening those 3D masks across the velocity dimension. We then used these bi-dimensional masks to measure the structures length. In order to compensate for low S/N  pixels that were masked in the previous step, we homogenised the bi-dimensional mask using {\sc scipy binary\_fill\_holes} task\footnote{\url{https://docs.scipy.org/doc/scipy-0.14.0/reference/generated/scipy.ndimage.morphology.binary_fill_holes.html}}. Applying the medial axis analysis implemented in the FilFinder2D package\footnote{\url{https://fil-finder.readthedocs.io/en/latest/}} \citep{koch2015}, we found the longest path across the skeleton of the structures, which constitute their spine. The length of the spine is equivalent to the length of the structure. The medial axis analysis works by labelling each pixel in the bi-dimensional projected mask with a number based on the distance (in pixels) to the structure boundary. Pixels within the mask with the largest distance from the boundary form the skeleton of the structure. The structure width was calculated as $2\times$ the median of the distance (in pixels) from the spine to the outer edge of the structure bi-dimensional mask (following \citealt{duarte-cabral2021}). For consistency, we also calculated the width as a median of the lengths of the spine `perpendicular cuts' implemented in the RadFil package\footnote{\url{https://github.com/catherinezucker/radfil}}. We found that our method tends to slightly underestimate the width of the structures compared to RadFil; however, the two measurements are largely comparable (see Table~\ref{T:catalogue} in Appendix~\ref{A:catalog}). The kinematic heliocentric distance to the structures was measured using the \cite{brand_blitz1993} Galaxy rotation curve model using $V_0=240$\,km\,s$^{-1}$ from \cite{reid2014}, which is consistent (within the uncertainties) with the updated value from \cite{reid2019}, $V_0=236\pm7$\,km\,s$^{-1}$. Also, we assumed $R_0=8.15$\,kpc \citep{reid2019}.



\subsection{Mass estimates}\label{A:mass}
The mass of a filament or generally of a molecular gas structure can be derived in a variety of ways, each of them presenting its own biases and approximations. For example, masses derived from dust continuum, such as in the case of the inner Galaxy filaments of \cite{zucker2018}, might be affected by foreground and background subtraction uncertainties. \cite{mattern2018a} and \cite{zhang2019} used the $^{13}$CO to derive the molecular gas mass. The $^{13}$CO transition is sensitive to a denser medium than the $^{12}$CO transition (implied in our study), and it is less affected by optical depth effects. In turn, $^{13}$CO suffers from more severe lower beam filling factor effects than $^{12}$CO. In essence, all methods aim to measure the H$_2$ column density, $N({\rm H_2})$, from which the molecular gas mass is calculated as follows: 

\begin{equation}\label{E:mass}
M = \mu_{\rm H_2} m_{\rm p} \sum_{lbv} N(\mathrm{H_2}) \Delta x\Delta y
,\end{equation}

where the mean molecular weight of the molecular hydrogen is $\mu_{\rm H_2}=2.8$, $m_{\rm p}$ is the proton mass (i.e. the mass of the hydrogen atom), and $\Delta x$ and $\Delta y$ represent the pixel size. \cite{benedettini2021} used three methods to derive the mass of the clouds in the outer Galaxy (for $220<l<240$ deg) for the Forgotten Quadrant Survey (FQS), finding that they all give very similar mass estimates. They calculated $N({\rm H_2})$ from the $^{12}$CO(1-0) emission, from the $^{13}$CO(1-0) column density, and from the far-infrared dust emission. This study is particularly relevant for our purposes since the FQS observed the molecular medium of the outer Milky Way in a Galactic region\ that will also be covered by the full OGHReS data once the survey is completed.  

For our OGLSFs, we chose to calculate the molecular gas mass from $^{12}$CO(2-1) intensity assuming a $^{12}$CO-to-H$_2$ conversion factor $X_{\rm CO}$. This method provides the H$_2$ column density as follows:

\begin{equation}\label{E:nh2_xco}
N(\mathrm{H_2}) = X_{\rm CO(1-0)}I_{\rm CO(2-1)}R_{21}^{-1},
\end{equation}

where $I_{\rm CO(2-1)}$ is the integrated intensity of the $^{12}$CO(2-1) emission. We assumed $X_{\rm CO(1-0)} = 2.3 \times 10^{20}$\,cm$^{-2}$\,(K\,km\,s$^{-1}$)$^{-1}$ inferred by \cite{brand_wouterloot1995} \citep[see also][]{heyer2001,bolatto2013,koenig2021} who showed that in the outer Galaxy, the CO-to-H$_2$ conversion factor, $X_{\rm CO(1-0)}$, is within the uncertainties consistent with
the inner Galaxy value. As $X_{\rm CO(1-0)}$ was calculated for the $^{12}$CO(1-0) line, we used a ration between the $^{12}$CO(2-1) to the $^{12}$CO(1-0) intensity, $R_{21}=0.7$ \citep[e.g. ][]{sandstrom2013,nishimura2015}. 

Modern studies, based on high resolution surveys, have shown that $X_{\rm CO}$ is not constant, by it varies pixel-by-pixel \citep[e.g ][]{pitts_barnes2021}, and it diverges from the assumed value especially at low integrated intensity. Therefore, here we calculate $N({\rm H_2})$ from the $^{13}$CO column density. Unfortunately, for our sources, the $^{13}$CO emission was only detected across the brightest regions within the filaments and not for every filament. Therefore, $^{13}$CO emission cannot be used to infer the whole mass of the OGLSFs, identified through the more extended and brighter $^{12}$CO emission, but only to test the reliability of our masses estimated through the $^{12}$CO emission. To do so, we first isolated the high S/N $^{13}$CO emission using a dilate masking technique (see Section~\ref{AA:method}) considering S/N=2,3 for the low and high masking thresholds, respectively. We then assumed that the molecular gas can be described as a system in local thermodymanic equilibrium (LTE) and that the $^{12}$CO(2-1) is optically thick. According to the formalism by \cite{wilson2013}, the excitation temperature, $T_{\rm ex}$, at every voxel ($l,b,v$) can be derived as follows:

\begin{equation}\label{E:Tex}
T_{\rm ex}\,[K] = \frac{11.06}{\ln[1 + 11.06/(T_{\rm mb}(^{12}\mathrm{CO}) + 0.194)]},    
\end{equation}

where $T_{\rm mb}(^{12}\mathrm{CO}$ is the main-beam brightness temperature of the $^{12}\mathrm{CO}$ emission at a given $(l,b,v)$ voxel. This equation is derived from equation 15.29 from \cite{wilson2013} considering a cosmic background temperature $T_{\rm bg}=2.725$\,K and $\nu_{\rm ^{12}CO(2-1)}=230.538$\,GHz.

The optical depth of the $^{13}$CO(2-1) is calculated at every $(l,b,v)$ voxel, solving equation 15.29 from \cite{wilson2013} for $\tau$:


\begin{equation}\label{E:tau_13}
\tau_{13} = -\ln\left[1-\frac{T_{\rm mb}(^{13}\mathrm{CO})/10.58}{(\exp{(10.58/T_{\rm ex})} -1)^{-1} - 0.02}\right],    
\end{equation}

where $T_{\rm mb}(^{13}\mathrm{CO}$ is the main-beam brightness temperature of the $^{13}\mathrm{CO}$ emission at a given $(l,b,v)$ voxel.

We derived the total $^{13}$CO column density following \cite{wilson2013}, equation 15.37:

\begin{equation}\label{E:N13co}
N(^{13}\mathrm{CO})\,[\mathrm{cm^{-2}}] = 1.5\times10^4\int\frac{T_{\rm ex}\exp(5.3/T_{\rm ex})}{1-\exp(-10.6/T_{\rm ex})}\tau_{13}\mathrm{d}v,
\end{equation}

where the velocity, $v$, is in km\,s$^{-1}$, and the integration runs across the full spectral extent of the structure. The column density of the molecular gas follows as:

\begin{equation}\label{E:nh2_n13co}
N(\mathrm{H_2})\,[\mathrm{cm^{-2}}] = \frac{[\mathrm{H_2}]}{[\mathrm{^{13}CO}]}N({\rm ^{13}CO}),
\end{equation}

where [H$_2$]/[$^{13}$CO] = $7.1\times10^5$ is the isotopic abundance ratio of the $^{13}$CO isotopologue relative to H$_2$ \citep{frerking1982}.

A further alternative is to infer $N(\mathrm{H_2})$  from the $^{13}$CO(2-1) integrated emission directly, by assuming a constant $X_\mathrm{^{13}CO(2-1)}$ factor. Such a $X-$factor for the $^{13}$CO(2-1) line has been calculated from SEDIGISM survey data \citep{schuller2017}. On average, the $X_\mathrm{^{13}CO(2-1)}$ they measured appears roughly constant with the intensity (even if a large scatter is observed). \cite{schuller2017} calculated the $X-$factor in two ways, comparing Hi-GAL dust continuum emission and $^{13}$CO(2-1) emission, and via a radiative transfer solution involving Three-mm Ultimate Mopra Milky Way Survey (ThrUMMS) data \citep{barnes2015}. Both methods infer an average $X-$factor consistent with $X_\mathrm{^{13}CO(2-1)}=1\times10^{21}$\,cm$^{-2}$\,(K\,km\,s$^{-1}$)$^{-1}$. Assuming that $X_\mathrm{^{13}CO(2-1)}$ does not present a significant variation with respect to the Galactocentric radius (such as $X_\mathrm{CO(1-0)}$), we can calculate $N(\mathrm{H_2})$ as follows:

\begin{equation}\label{E:nh2_x13co}
N(\mathrm{H_2})\,[\mathrm{cm^{-2}}] = X_{\rm ^{13}CO(2-1)}I_{\rm ^{13}CO(2-1)}.
\end{equation}

We named the mass estimates as $M_{X(^{12}\mathrm{CO})}$, $M_{N(\mathrm{^{13}CO})}$, and $M_{X(^{13}\mathrm{CO})}$, considering $N(\mathrm{H_2})$ inferred from equations~\ref{E:nh2_xco},~\ref{E:nh2_n13co}, and ~\ref{E:nh2_x13co}, respectively. To perform this test, $M_{X(^{12}\mathrm{CO})}$ was calculated across the same area defined by the $^{13}$CO(2-1) emission mask for each filament in our sample. 
\begin{table}
\setlength{\tabcolsep}{3.2pt}
\centering
\caption{Ratio of the masses of the clumps within the OGLSFs derived using various methods}.
\begin{tabular}{cccc}
\hline
Mass ratio & Median & 25$^{th}$ percentile & 75$^{th}$ percentile \\
\hline
$M_{X(^{12}\mathrm{CO})}$/$M_{N(\mathrm{^{13}CO})}$ & 1.08 & 0.83 & 1.49 \\
$M_{X(^{12}\mathrm{CO})}$/$M_{X(^{13}\mathrm{CO})}$ & 1.00 & 0.88 & 1.14 \\
\hline
\end{tabular}
\label{T:mass_ratios}
\tablefoot{See description in Section~\ref{A:mass}.}
\end{table}

Medians and percentile values for the mass ratio distributions calculated through the three methods discussed here are summarised in Table~\ref{T:mass_ratios}. Generally, the masses estimated from the $^{13}$CO emission and column density largely agree with the masses inferred from the $^{12}$CO emission assuming a constant $X-$factor. The most significant variations are observed for filaments where the $^{13}$CO emission is barely detected. Given this, we expect the masses derived for the whole OGLSFs through the $^{12}$CO emission to be a reliable estimation of the actual molecular gas mass of the structures.


However, looking at Fig.~\ref{F:props}, it appears that different mass calculation methods give different estimates for the filaments in the inner Galaxy. LSF masses from \cite{zhang2019} appear to be a factor of a few larger than the masses measured by \cite{zucker2018} (see Fig.~\ref{F:props}). On average, however, the lengths are more comparable. This results in larger linear masses than the ones inferred by \cite{zucker2018} (further increasing the difference between inner and outer Galaxy LSF linear masses). \cite{zucker2018} used HI-GAL data to trace the filament boundaries (and infer their masses from $N({\rm H_2})$ inferred from dust continuum), in an attempt to maintain the structure shape and appearance compared to their original studies. In turn, \cite{zhang2019} measured the LSF properties from $^{13}$CO(1-0) data (including their mass through $N(^{13}\mathrm{CO})$), assuming a constant $V-$band extinction, $A_{\rm v}=3$, to define the boundary of each structure in their sample. In this aspect, \cite{zucker2018} consider the denser part of the filaments, resulting in lower masses and linear masses. Therefore, discrepancies between the measurements of \cite{zucker2018} and \cite{zhang2019} are possibly due to the different methods used to define the LSFs, rather than inconsistencies between dust and $^{13}$CO emission to calculate the structure masses. Indeed, for the GMFs, for which both works used the structure masks from the original studies, mass distributions appear more comparable.

\section{Catalogue of properties and integrated intensity map atlas of the outer Galaxy large-scale filaments}\label{A:catalog}
\begin{figure*}
\centering
\includegraphics[width=0.9\textwidth]{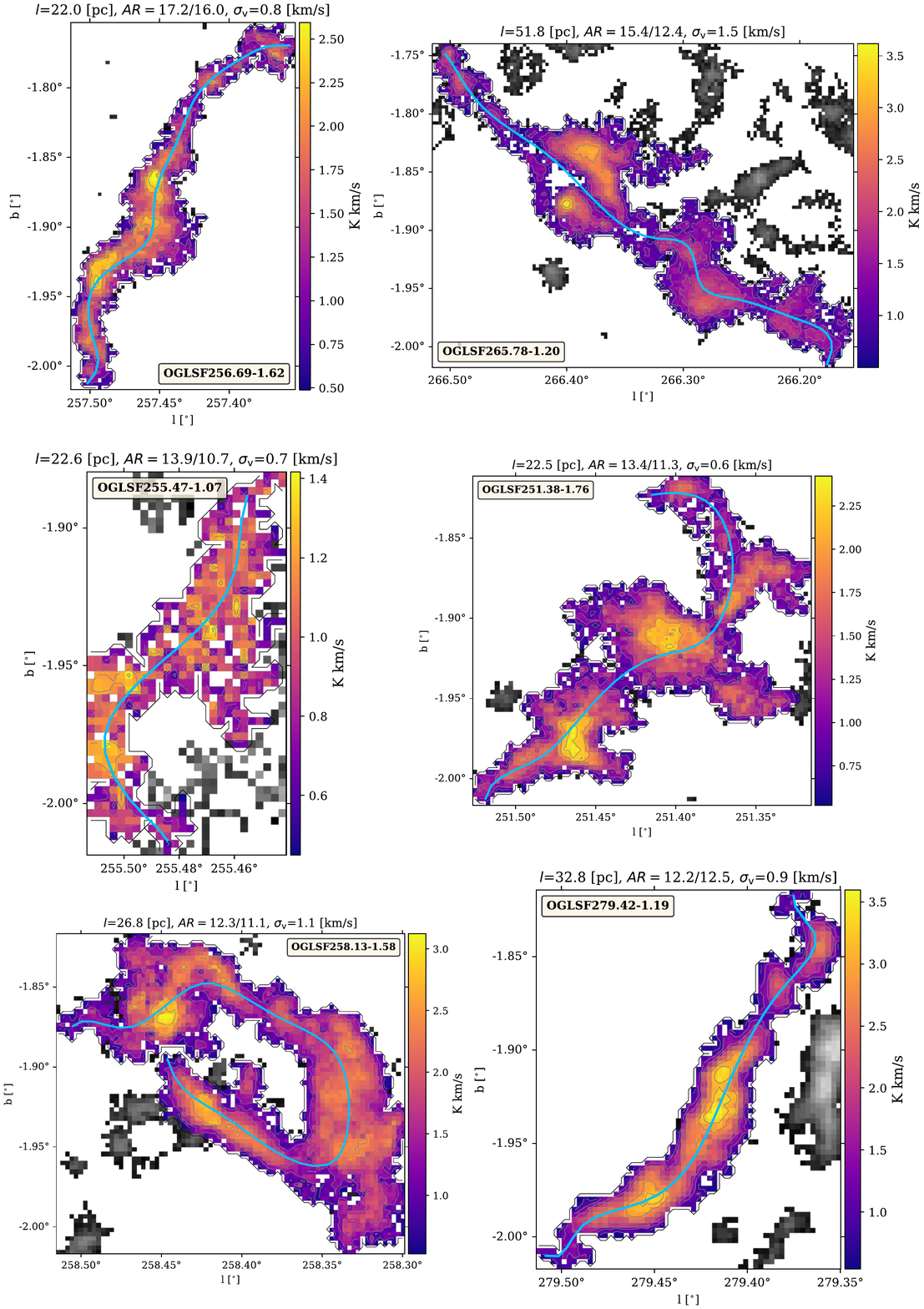}
\caption{Integrated intensity maps of the $^{12}$CO(2-1) emission from OGLSFs. Symbols and conventions are the same as in Fig.~\ref{F:oglsf_mom0}.}
\label{F:oglsf_mom0_01}
\end{figure*}

\begin{figure*}
\centering
\includegraphics[width=0.9\textwidth]{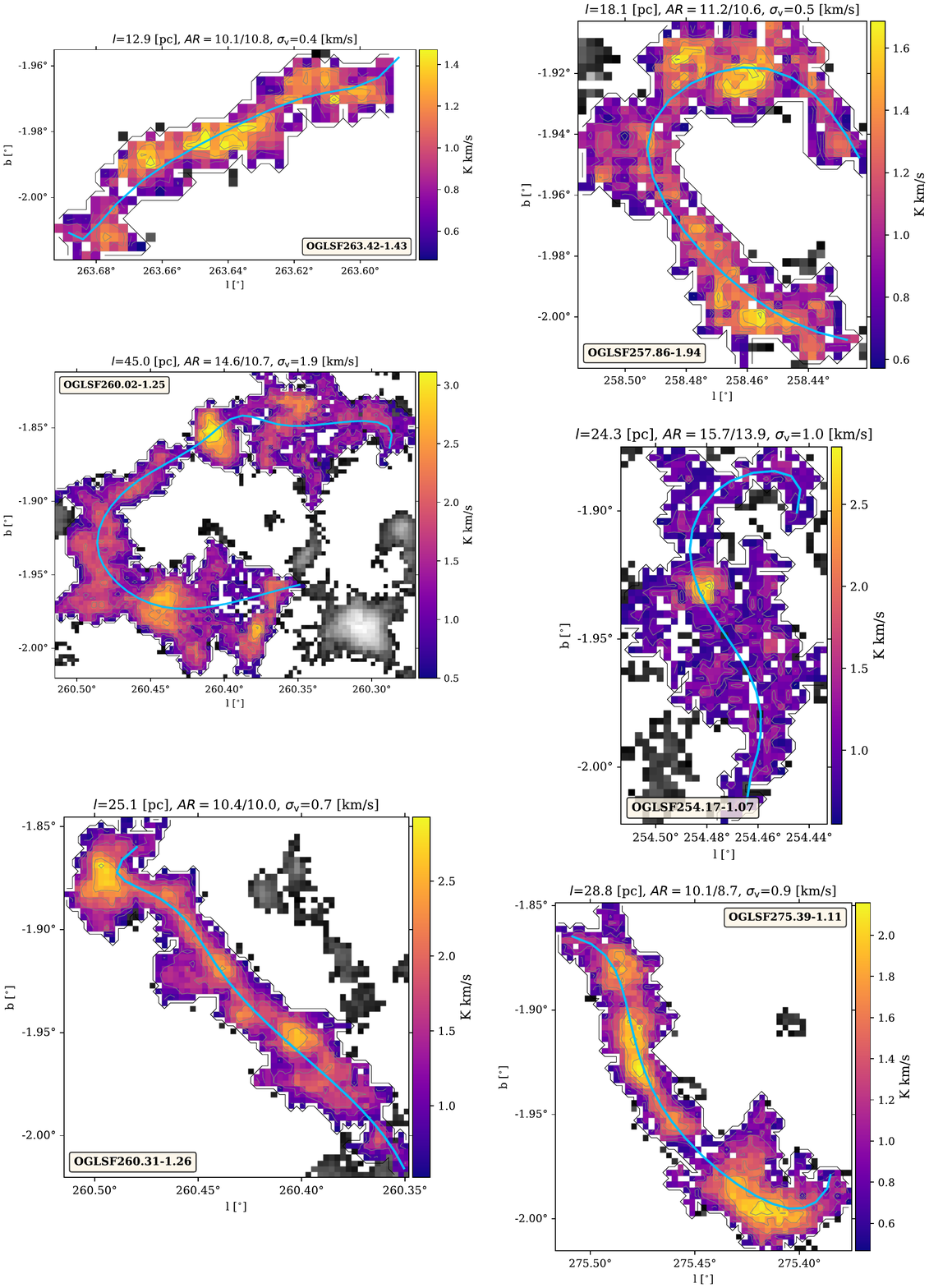}
\caption{Integrated intensity maps of the $^{12}$CO(2-1) emission from OGLSFs. Symbols and conventions are the same as in Fig.~\ref{F:oglsf_mom0}.}
\label{F:oglsf_mom0_02}
\end{figure*}

\begin{figure*}
\centering
\includegraphics[width=0.9\textwidth]{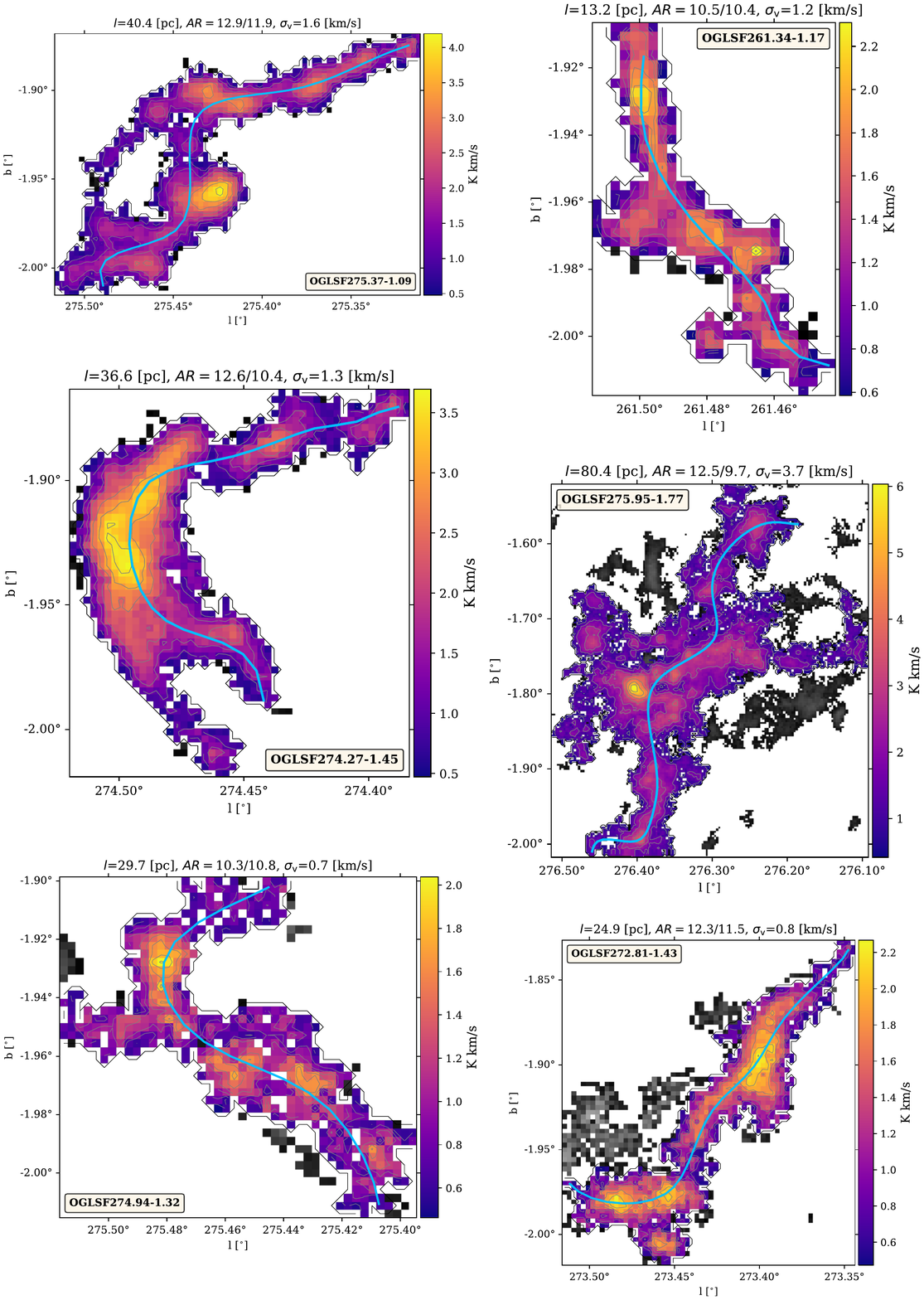}
\caption{Integrated intensity maps of the $^{12}$CO(2-1) emission from OGLSFs. Symbols and conventions  are the same as in Fig.~\ref{F:oglsf_mom0}.}
\label{F:oglsf_mom0_03}
\end{figure*}

\begin{figure*}
\centering
\includegraphics[width=0.9\textwidth]{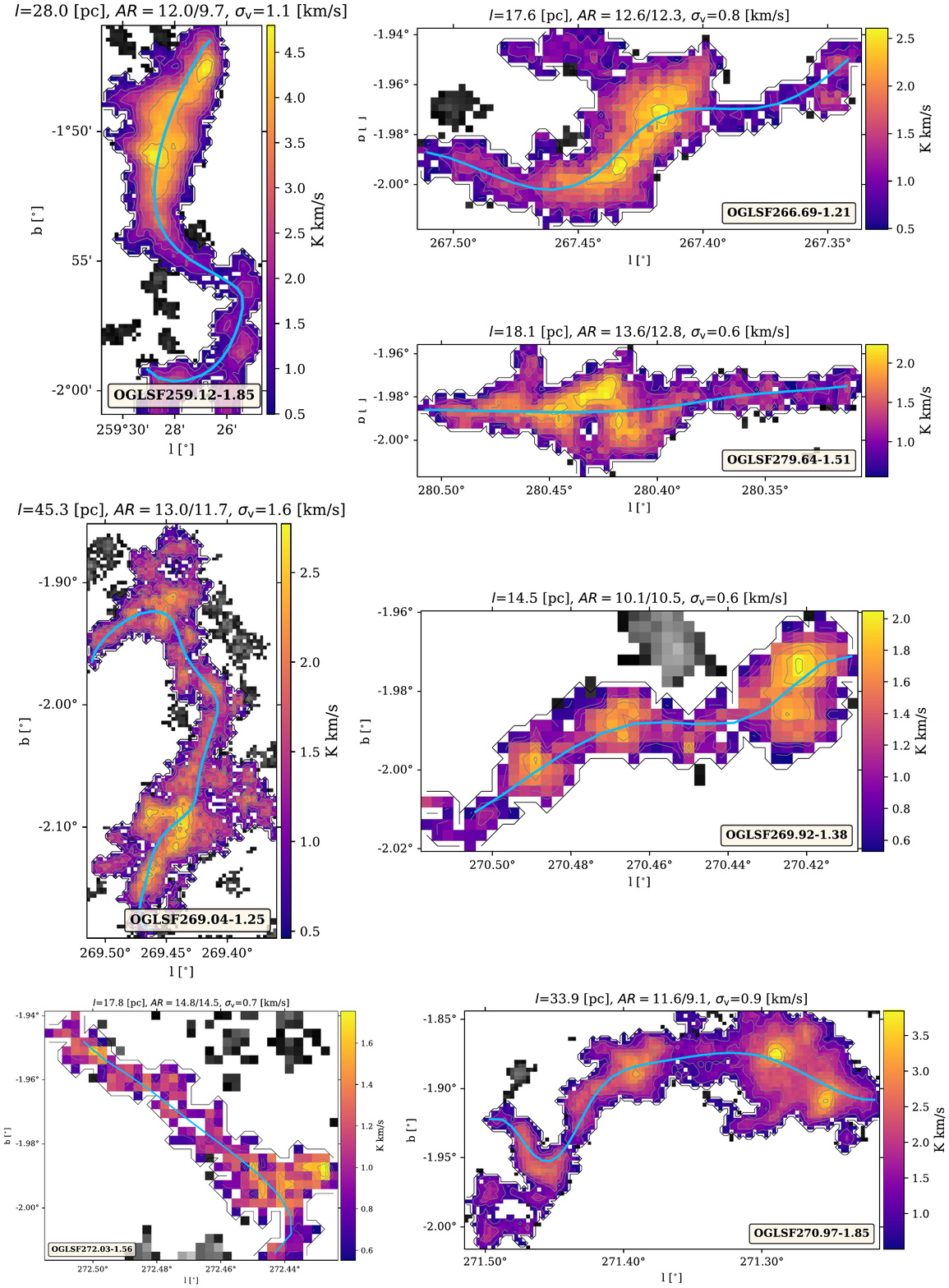}
\caption{Integrated intensity maps of the $^{12}$CO(2-1) emission from OGLSFs. Symbols and conventions  are the same as in Fig.~\ref{F:oglsf_mom0}.}
\label{F:oglsf_mom0_04}
\end{figure*}

\begin{figure*}
\centering
\includegraphics[width=0.9\textwidth]{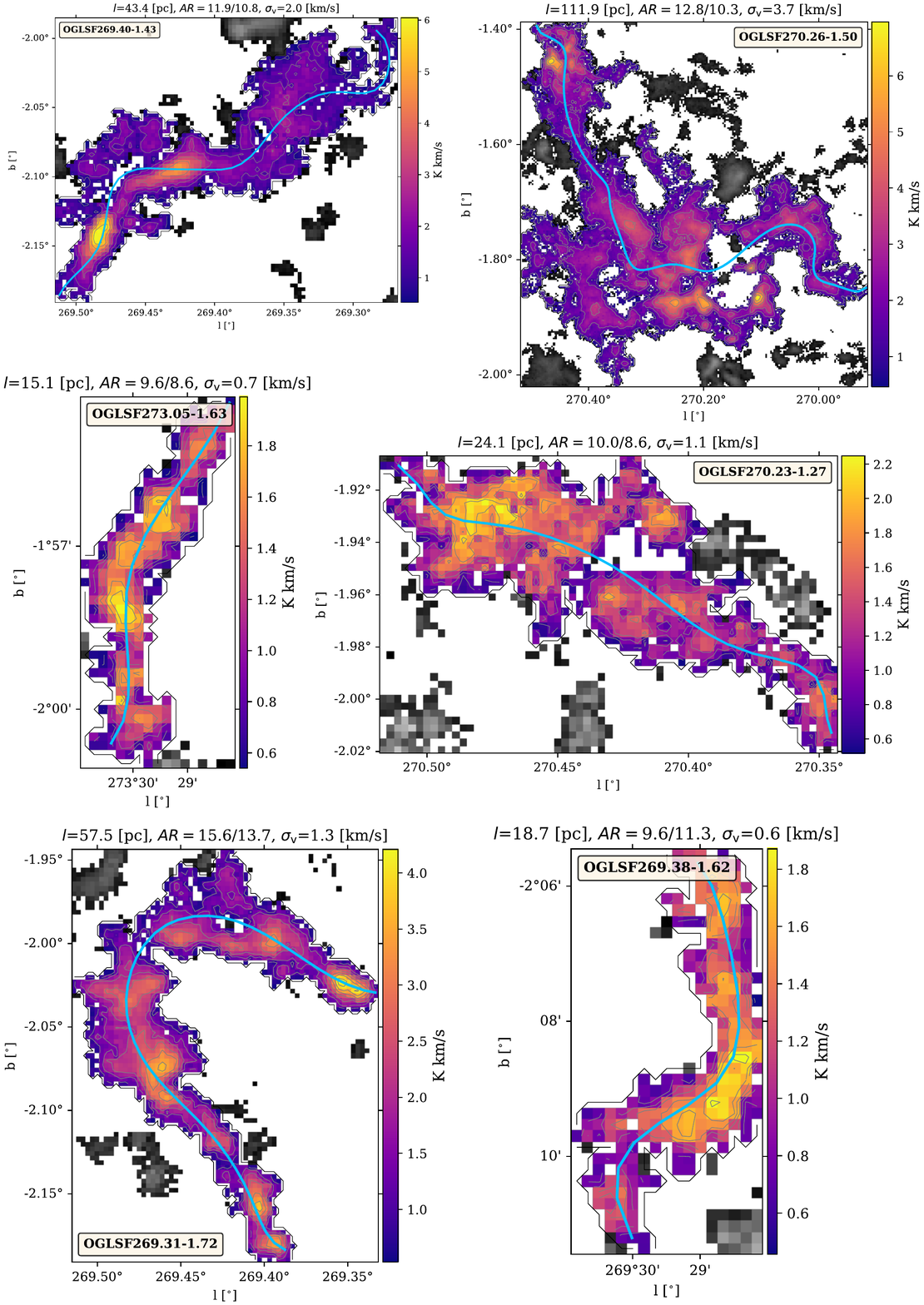}
\caption{Integrated intensity maps of the $^{12}$CO(2-1) emission from OGLSFs. Symbols and conventions  are the same as in Fig.~\ref{F:oglsf_mom0}.}
\label{F:oglsf_mom0_05}
\end{figure*}

Here, we provide in Table~\ref{T:catalogue} the catalogue of the OGLSF properties discussed in the paper. Together we show the remaining OGLSF integrated intensity maps (Fig.~\ref{F:oglsf_mom0_01}-\ref{F:oglsf_mom0_05}). Sub-cubes containing the $^{12}$CO(2-1) emission from the 37 OGLSFs can be obtained from CDS.

\begin{table*}
\setlength{\tabcolsep}{3.2pt}
\renewcommand{\arraystretch}{1.2}
\centering
\caption{Catalogue of the OGLSFs identified within the OGHReS SD1 field}
\begin{tabular}{cccccccccccccc}
\hline
Name & $l$ & $b$ & $V_{\rm LSR}$ & $D$ & $x_{\rm Gal}$ & $y_{\rm Gal}$ & $\sigma_{\rm v}$ & $V_{\rm 5th}/V_{\rm 95th}$ & $\log(M_{\rm CO})$ & \emph{length} & $AR_{\rm MA}$ & $AR_{\rm RadFil}$ & $\Delta V_{\rm sp. arm}$ \\
 & $\mathrm{{}^{\circ}}$ & $\mathrm{{}^{\circ}}$ & $\mathrm{km\,s^{-1}}$ & kpc & kpc & kpc & $\mathrm{km\,s^{-1}}$ & $\mathrm{km\,s^{-1}}$ & $M_{\odot}$ & $\mathrm{pc}$ &  &  & $\mathrm{km\,s^{-1}}$ \\
\hline
OGLSF251.38-1.76 & 251.38 & -1.76 & 49.2 & 4.34 & -4.11 & 9.49 & 0.6 & 48/50 & 3.00 & 22.5 & 13.4 & 11.3 & 27.9 \\
OGLSF254.17-1.07 & 254.17 & -1.07 & 77.3 & 7.15 & -6.88 & 10.05 & 1.0 & 75/79 & 2.79 & 24.3 & 15.7 & 13.9 & 2.5 \\
OGLSF255.47-1.07 & 255.47 & -1.07 & 80.2 & 7.53 & -7.29 & 9.99 & 0.7 & 78/81 & 2.60 & 22.6 & 13.9 & 10.7 & 0.8 \\
OGLSF256.69-1.62 & 256.69 & -1.62 & 42.8 & 4.16 & -4.04 & 9.06 & 0.8 & 41/44 & 2.82 & 22.0 & 17.2 & 16.0 & 39.2 \\
OGLSF257.86-1.94 & 257.86 & -1.94 & 57.8 & 5.57 & -5.44 & 9.27 & 0.5 & 57/58 & 2.53 & 18.1 & 11.2 & 10.6 & 25.3 \\
OGLSF258.13-1.58 & 258.13 & -1.58 & 41.5 & 4.16 & -4.07 & 8.96 & 1.1 & 39/43 & 3.26 & 26.8 & 12.3 & 11.1 & 41.8 \\
OGLSF259.12-1.85 & 259.12 & -1.85 & 54.6 & 5.39 & -5.30 & 9.12 & 1.1 & 52/56 & 3.58 & 28.0 & 12.0 & 9.7 & 29.6 \\
OGLSF259.82-1.38 & 259.82 & -1.38 & 61.5 & 6.07 & -5.98 & 9.17 & 1.5 & 58/63 & 3.48 & 55.9 & 26.3 & 26.0 & 23.4 \\
OGLSF260.02-1.25 & 260.02 & -1.25 & 63.0 & 6.23 & -6.14 & 9.18 & 1.9 & 59/65 & 3.51 & 45.0 & 14.6 & 10.7 & 22.0 \\
OGLSF260.31-1.26 & 260.31 & -1.26 & 62.9 & 6.25 & -6.16 & 9.15 & 0.7 & 61/64 & 3.14 & 25.1 & 10.4 & 10.0 & 22.4 \\
OGLSF261.31-1.08 & 261.31 & -1.08 & 66.0 & 6.64 & -6.56 & 9.10 & 0.9 & 64/67 & 3.57 & 44.8 & 11.6 & 10.7 & 20.1 \\
OGLSF261.34-1.17 & 261.34 & -1.17 & 56.7 & 5.79 & -5.73 & 8.97 & 1.2 & 54/58 & 2.48 & 13.2 & 10.5 & 10.4 & 29.4 \\
OGLSF263.42-1.43 & 263.42 & -1.43 & 55.5 & 5.91 & -5.87 & 8.78 & 0.4 & 54/56 & 2.12 & 12.9 & 10.1 & 10.8 & 32.3 \\
OGLSF265.59-1.48 & 265.59 & -1.48 & 63.7 & 6.90 & -6.88 & 8.63 & 0.8 & 62/65 & 3.64 & 66.2 & 22.1 & 14.7 & 25.8 \\
OGLSF265.78-1.20 & 265.78 & -1.20 & 62.5 & 6.81 & -6.79 & 8.60 & 1.5 & 58/64 & 3.67 & 51.8 & 15.4 & 12.4 & 27.2 \\
OGLSF266.69-1.21 & 266.69 & -1.21 & 42.1 & 5.11 & -5.11 & 8.40 & 0.8 & 40/43 & 2.77 & 17.6 & 12.6 & 12.3 & 48.3 \\
OGLSF269.04-1.25 & 269.04 & -1.25 & 61.9 & 7.18 & -7.18 & 8.22 & 1.6 & 59/65 & 3.64 & 45.3 & 13.0 & 11.7 & 30.2 \\
OGLSF269.31-1.72 & 269.31 & -1.72 & 82.3 & 9.19 & -9.19 & 8.21 & 1.3 & 80/84 & 3.91 & 57.5 & 15.6 & 13.7 & 10.0 \\
OGLSF269.38-1.62 & 269.38 & -1.62 & 80.0 & 8.97 & -8.97 & 8.20 & 0.6 & 79/81 & 2.65 & 18.7 & 9.6 & 11.3 & 12.3 \\
OGLSF269.40-1.43 & 269.40 & -1.43 & 59.3 & 7.00 & -7.00 & 8.17 & 2.0 & 55/62 & 3.74 & 43.4 & 11.9 & 10.8 & 33.0 \\
OGLSF269.92-1.38 & 269.92 & -1.38 & 54.6 & 6.65 & -6.65 & 8.11 & 0.6 & 53/55 & 2.45 & 14.5 & 10.1 & 10.5 & 38.1 \\
OGLSF270.23-1.27 & 270.23 & -1.27 & 52.8 & 6.53 & -6.53 & 8.07 & 1.1 & 50/54 & 3.03 & 24.1 & 10.0 & 8.6 & 40.1 \\
OGLSF270.26-1.50 & 270.26 & -1.50 & 54.7 & 6.71 & -6.71 & 8.07 & 3.7 & 48/59 & 4.81 & 111.9 & 12.8 & 10.3 & 38.2 \\
OGLSF270.97-1.85 & 270.97 & -1.85 & 44.8 & 5.93 & -5.93 & 8.00 & 0.9 & 43/46 & 3.59 & 33.9 & 11.6 & 9.1 & 48.6 \\
OGLSF272.03-1.56 & 272.03 & -1.56 & 73.7 & 8.72 & -8.72 & 7.79 & 0.7 & 72/74 & 2.43 & 17.8 & 14.8 & 14.5 & 20.4 \\
OGLSF272.81-1.43 & 272.81 & -1.43 & 39.9 & 5.78 & -5.77 & 7.82 & 0.8 & 38/41 & 2.94 & 24.9 & 12.3 & 11.5 & 54.7 \\
OGLSF273.05-1.63 & 273.05 & -1.63 & 55.8 & 7.22 & -7.21 & 7.72 & 0.7 & 54/57 & 2.54 & 15.1 & 9.6 & 8.6 & 38.9 \\
OGLSF274.07-1.20 & 274.07 & -1.20 & 38.5 & 5.86 & -5.84 & 7.68 & 2.4 & 34/42 & 4.82 & 143.4 & 20.8 & 17.0 & 56.8 \\
OGLSF274.27-1.45 & 274.27 & -1.45 & 77.3 & 9.43 & -9.40 & 7.40 & 1.3 & 75/79 & 3.67 & 36.6 & 12.6 & 10.4 & 18.1 \\
OGLSF274.94-1.32 & 274.94 & -1.32 & 87.1 & 10.56 & -10.52 & 7.19 & 0.7 & 86/88 & 3.00 & 29.7 & 10.3 & 10.8 & 8.7 \\
OGLSF275.32-1.69 & 275.32 & -1.69 & 56.4 & 7.63 & -7.60 & 7.39 & 1.1 & 54/58 & 3.87 & 57.3 & 11.0 & 9.5 & 39.7 \\
OGLSF275.37-1.09 & 275.37 & -1.09 & 70.7 & 8.96 & -8.92 & 7.26 & 1.6 & 68/73 & 3.66 & 40.4 & 12.9 & 11.9 & 25.4 \\
OGLSF275.39-1.11 & 275.39 & -1.11 & 61.9 & 8.14 & -8.10 & 7.33 & 0.9 & 60/63 & 3.15 & 28.8 & 10.1 & 8.7 & 34.2 \\
OGLSF275.95-1.77 & 275.95 & -1.77 & 56.6 & 7.75 & -7.71 & 7.30 & 3.7 & 51/63 & 4.46 & 80.4 & 12.5 & 9.7 & 39.8 \\
OGLSF278.85-1.33 & 278.85 & -1.33 & 45.8 & 7.31 & -7.22 & 6.97 & 1.2 & 43/48 & 4.10 & 64.1 & 14.9 & 11.0 & 52.0 \\
OGLSF279.42-1.19 & 279.42 & -1.19 & 39.9 & 6.90 & -6.80 & 6.97 & 0.9 & 38/41 & 3.41 & 32.8 & 12.2 & 12.5 & 58.1 \\
OGLSF279.64-1.51 & 279.64 & -1.51 & 16.8 & 4.85 & -4.78 & 7.29 & 0.6 & 16/18 & 2.63 & 18.1 & 13.6 & 12.8 & 81.4 \\
\hline
\end{tabular}
\tablefoot{\emph{From left to right:} name of the source (Name); Galactic longitude ($l$); Galactic latitude ($b$); LSR velocity ($V_{\rm LSR}$); heliocentric distance ($D$); Galactocentric $x$ coordinate ($x_{\rm Gal}$); Galactocentric $y$ coordinate ($y_{\rm Gal}$); velocity dispersion ($\sigma_{\rm v}$); fifth and 95th percentile of the velocity distribution span by a filament ($V_{\rm 5th}$ and $V_{\rm 95th}$, respectively); logarithmic mass from the CO luminosity ($M_{\rm CO}$); filament length (\emph{length}); aspect ratio with the width inferred from the medial axis analysis ($AR_{\rm MA}$); aspect ratio with the width inferred from RadFil ($AR_{\rm RadFil}$); and velocity offset with respect to Perseus spiral arm considering \cite{taylor_cordes1993} models ($\Delta V_{\rm sp. arm}$).}
\label{T:catalogue}
\end{table*}

\clearpage
\newpage

\end{document}